\documentclass{article}
\usepackage{amsmath,amssymb,amsthm}

\usepackage[width=0.95\textwidth,indention=0mm,justification=centerlast,
 font=small,labelfont=bf,labelsep=period]{caption}

\usepackage[pdfpagemode=UseOutlines,
bookmarks=true, bookmarksopenlevel=3, bookmarksopen=true,
bookmarksnumbered=true, unicode=true,
pdffitwindow=true,pdfpagelayout=SinglePage,pdfhighlight=/P,
colorlinks=true,linkcolor=blue,citecolor=blue,urlcolor=blue]{hyperref}

\advance\textwidth 20mm  \advance\oddsidemargin -10mm
\advance\textheight 20mm \advance\topmargin -15mm
\allowdisplaybreaks[3]

\theoremstyle{plain}
\newtheorem{theorem}{Theorem}

\theoremstyle{remark}
\newtheorem{remark}{Remark}
\newtheorem{example}{Example}

\title{Non-Abelian evolution systems with conservation laws}

\date{20 August 2020}

\author{V.E. Adler\thanks{L.D.~Landau Institute for Theoretical Physics, Chernogolovka, Russian Federation.\newline E-mail:~adler@itp.ac.ru},~ V.V. Sokolov$^*$\thanks{Federal University of ABC, Santo Andr\'e, Sao Paulo, Brazil.}}

\begin{document}
\maketitle

\begin{abstract}
We find noncommutative analogs for well-known polynomial evolution systems with higher conservation laws and symmetries. The integrability of obtained non-Abelian systems is justified by explicit zero curvature representations with spectral parameter. 
\medskip

\noindent{\small Keywords: non-Abelian system, conservation law, symmetry, zero curvature representation}
\end{abstract}

%-------------------------------------------------------------------------------
\section{Introduction}

\subsection{Polynomial homogeneous evolution equations}

The right-hand side of most popular nonlinear evolution equations integrable by the inverse scattering method \cite{Ablowitz_Segur_1981} is a homogeneous differential polynomial. We say that the differential equation
\begin{equation}\label{eveq}
 u_t=F(u,u_x, u_{xx},\dots, u_n), \qquad u_i=\frac{\partial^i u}{\partial x^i},
\end{equation}
is {\em homogeneous with weights $\mu$ and $\nu$} if it admits the one-parameter scaling group 
\[
 (x, \ t, \ u)\longrightarrow (\tau^{-1}x, \ \tau^{-\mu} t, \ \tau^{\nu} u).
\]
For an $N$-component system with unknowns $u^1,\dots,u^N$, the scaling group is of the form
\begin{equation}
 \label{homo}(x,t,u^1,\dots,u^N)\longrightarrow 
 (\tau^{-1} x, \ \tau^{-\mu} t, \ \tau^{\nu_1} u^1,\dots, \tau^{\nu_N} u^N).
\end{equation}
For instance, the celebrated Korteweg--de Vries equation
\begin{equation} \label{kdv} 
u_t=u_{xxx}+ 6\,u\,u_x
\end{equation}
is homogeneous with the weights $\mu =3$ and $\nu =2$, the modified KdV equation
\begin{equation}\label{mkdv}
u_t=u_{xxx}-6 u^2 u_x
\end{equation}
is homogeneous with $\mu = 3$ and $\nu =1$, and for the nonlinear Schr\"odinger equation written as the coupled system
\begin{equation}\label{NLS}
u_t=-u_{xx}+2 u^2 v, \qquad v_t=v_{xx}-2 v^2 u,
\end{equation}
the weights can be chosen as $\mu=2$, $\nu_1=\varepsilon$ and $\nu_2 = 2 - \varepsilon$, where $\varepsilon$ is an arbitrary parameter.

It turns out that if the weight $\nu$ of an integrable homogeneous polynomial equation (\ref{eveq}) is positive then it can be equal only to the values $\nu=2$, $\nu=1$ or $\nu=\frac{1}{2}$ \cite{sw}. There is no such description for systems of several equations.

\subsection{Symmetries and conservation laws}

Both equations \eqref{kdv} and \eqref{NLS} possess higher infinitesimal symmetries (or higher flows) \cite{Olv93} and higher conservation laws. Recall that a higher symmetry is an evolution equation
\begin{equation}\label{evsym}
 u_{\tau}=G(u, u_x, u_{xx}, \dots , u_m), \qquad m > 1,
\end{equation}
which is consistent with \eqref{eveq}. The simplest higher symmetry for the KdV equation is of the form
\begin{equation}\label{kdvsym}
u_{\tau}=u_{xxxxx}+10 u u_{xxx}+20 u_x u_{xx}+30 u^2 u_x.
\end{equation}

A local conservation law for equation \eqref{eveq} is a relation of the form
\begin{equation}\label{rhosig}
 \bigl(\rho(u, u_x, u_{xx}, \dots)\bigr)_t = \bigl(\sigma(u, u_x, u_{xx}, \dots)\bigr)_x,
\end{equation}
where the $t$-derivative on the left-hand side is calculated in virtue of the evolution equation \eqref{eveq}. For the polynomial homogeneous equations, the functions $\rho$ and $\sigma$ are homogeneous differential polynomials. The function $\rho$ is called the density of conservation law. Few simplest densities for the KdV equation are 
\[
 \rho_1=u,\qquad \rho_2=u^2,\qquad \rho_3=-u_x^2+2u^3. 
\]
It is well-known that \eqref{rhosig} implies the relation
\begin{equation}\label{varrho}
 \frac{\delta}{\delta u}(\rho_t) = 0,
\end{equation}
where
\[
 \frac{\delta}{\delta u}=\sum_k (-1)^k D^k \circ \frac{\partial}{\partial u_k}
\]
is the {\em Euler operator} or the {\em variational derivative}. Here and below, $D$ denotes the total $x$-derivative
\begin{equation}\label{DxShort}
 D = \sum_{i=0}^\infty u_{i+1} \frac{\partial}{\partial u_i}.
\end{equation}
The relation \eqref{rhosig} can be written as $\rho_t \in \operatorname{Im}D$. A density $\rho$ of a conservation law is defined up to addition of total $x$-derivatives. In other words, $\rho$ is an element of the quotient space of the algebra of differential polynomials over the subspace $\operatorname{Im}D$.

The existence of higher symmetries was adopted as the basis of the symmetry approach to the classification of integrable evolution equations and coupled systems \cite{MikShaSok91, ASY}. Another, related and more stringent criterion for integrability is associated with the existence of higher  conservation laws for \eqref{eveq}.

\subsection{Non-Abelian evolution equations}

It is well-known \cite{march, etgelret, kuper, SviSok94, OS_1998a} that many integrable equations and their symmetries admit matrix generalizations. For instance, the matrix KdV equation
\begin{equation}\label{matkdv}
 {\bf u}_t = {\bf u}_{xxx} + 3{\bf u} {\bf u}_x + 3 {\bf u}_x {\bf u},
\end{equation}
where ${\bf u}(x,t)$ is an unknown $m\times m$ matrix, has infinitely many matrix symmetries for any $m$. The simplest one is of the form
\begin{equation}\label{matkdvsym}
 {\bf u}_{\tau}={\bf u}_{xxxxx}+5\, ({\bf u} {\bf u}_{xxx}+{\bf u}_{xxx} {\bf u})+10\,({\bf u}_{x} {\bf u}_{xx}+{\bf u}_{xx} {\bf u}_{x})
  +10\,({\bf u}^{2} {\bf u}_{x}+{\bf u} {\bf u}_{x} {\bf u}+{\bf u}_{x}{\bf u}^{2}).
\end{equation}
If $m=1$ then equation \eqref{matkdv} coincides with \eqref{kdv} and \eqref{matkdvsym} coincides with \eqref{kdvsym}. The mKdV equation \eqref{mkdv} admits two different matrix generalizations (see e.g. \cite[sect.~3.9]{kuper})
\begin{equation}\label{matmkdv}
{\bf u}_t={\bf u}_{xxx}-3 {\bf u}^2 {\bf u}_x-3 {\bf u}_x {\bf u}^2
\end{equation}
and
\begin{equation}\label{mat2}
{\bf u}_{t}={\bf u}_{xxx}+3 {\bf u} {\bf u}_{xx}-3 {\bf u}_{xx} {\bf u}-6 {\bf u} {\bf u}_x {\bf u}.
\end{equation}
A matrix generalization of the NLS system \eqref{NLS} is of the form
\begin{equation}\label{matnls}
{\bf u}_t={\bf u}_{xx}-2\, {\bf u} {\bf v} {\bf u}, \qquad {\bf v}_t=-{\bf v}_{xx}+2\, {\bf v} {\bf u} {\bf v}.
\end{equation} 
 
In calculations related to matrix equations, we always work not with matrix elements, but with noncommutative associative polynomials. Therefore, it is convenient to adopt a formalized algebraic point of view on matrix evolution equations like \eqref{matnls}, which treats the variables ${\bf u},{\bf v},{\bf u}_x,{\bf v}_x,\dots$ as generators of a free associative algebra $\cal A$ over $\mathbb{C}$  \cite{int2,Sokolov}. In this language, an evolution {\em non-Abelian} equation is a derivation $D_t$ of the algebra $\cal A$ commuting with the derivation $D$, its symmetry is defined as a derivation $D_{\tau}$ such that $[D, D_{\tau}] = [D_t, D_{\tau}]=0$, and a (non-Abelian) density of a conservation law is an element $\rho$ of the linear quotient space ${\cal A}/{\cal T}$, where
\[
 {\cal T} = [{\cal A},{\cal A}]+\operatorname{Im}D,
\]
such that $D_t(\rho)\in {\cal T}$. Often, we denote by $\rho\in{\cal A}$ some representative of the corresponding equivalence class, in hope that this will not lead to a misunderstanding.

Less formally, the definition of conserved density in the matrix case means that the value of the functional
\begin{equation}\label{funk}
 I = \int_{-\infty}^{+\infty}\operatorname{trace}\rho({\bf u},{\bf v},{\bf u}_x,{\bf v}_x,\dots)\,dx
\end{equation}
does not depend on $t$ for solutions of the system which rapidly decrease for $x\to\pm\infty$. It is clear that the value of this functional does not change when the total $x$-derivative of a matrix polynomial or the commutator of matrix polynomials is added to $\rho$.

\section{Setting of the problem and brief description of results}
\label{s:com}

In papers \cite{Mikhailov_Shabat_1985,MikShaYam87}, a classification of coupled systems of the form
\begin{equation}\label{sys2}
 u_t=u_{xx}+F(u,v,u_x,v_x), \qquad v_t=-v_{xx}+G(u,v,u_x,v_x),
\end{equation}
admitting higher conservation laws was obtained. The found integrable systems can be subdivided into those that have a symmetry of order 3 and those that have no third order symmetry, but have a symmetry of order 4. We call the first group NLS type systems, after the system \eqref{NLS}. The conventional name for the second group is the Boussinesq type systems.

In the obtained list, many systems have a polynomial right-hand side. In particular, the following statement holds.

\begin{theorem}[\cite{MikShaYam87, SW}]\label{th:com} 
{\rm i)} If a system of the form
\begin{equation}\label{kvazgen}
\begin{cases}
 u_t = u_{xx} + A_{1}(u,v)\,u_x + A_{2}(u,v)\,v_x + A_{0}(u,v), \\[1.5mm]
 v_t = - v_{xx} + B_{1}(u,v)\,v_x + B_{2}(u,v)\,u_x + B_{0}(u,v)
\end{cases}
\end{equation}
admits an infinite sequence of conservation laws then it is polynomial.

{\rm ii)} A homogeneous polynomial non-triangular system \eqref{kvazgen} admits higher conservation laws if and only if it belongs to one of the following lists, up to the scaling of the variables $x,t,u,v$ and the interchange $(u,v)\mapsto(v,u)$: 
\begin{align}
\intertext{1. NLS type systems}
%--------------------------------------------- 
\label{S1}\tag*{\text{$S_1$}}
&\left\{\begin{aligned}
 u_t &=~~u_{xx}+2uvu,\\
 v_t &= -v_{xx}-2vuv,
\end{aligned}\right.\\
%--------------------------------------------- 
\label{S2}\tag*{\text{$S_2$}}
&\left\{\begin{aligned}
 u_t &=~~u_{xx }+2uu_{x}+2vu_{x}+2uv_{x},\\
 v_t &= -v_{xx }+2vv_{x}+2vu_{x}+2uv_{x},
\end{aligned}\right.\\
%--------------------------------------------- 
\label{S3}\tag*{\text{$S_3$}}
&\left\{\begin{aligned}
 u_t &=~~u_{xx }+2u_{x}v+2uv_{x},\\
 v_t &= -v_{xx }+2vv_{x}+2u_{x},
\end{aligned}\right.\\
%--------------------------------------------- 
\label{S4}\tag*{\text{$S_4$}} 
&\left\{\begin{aligned}
 u_t &=~~u_{xx}+2(u+v)u_{x},\\
 v_t &= -v_{xx}+2(u+v)v_{x},
\end{aligned}\right.\\
%--------------------------------------------- 
\label{S5}\tag*{\text{$S_5$}} 
&\left\{\begin{aligned}
 u_t &=~~u_{xx }+2\alpha u^2v_{x}+2\beta uvu_{x}+\alpha(\beta-2\alpha)u^3v^2,\\
 v_t &= -v_{xx }+2\alpha v^2u_{x}+2\beta uvv_{x}-\alpha(\beta-2\alpha)u^2v^3;
\end{aligned}\right.\\
\intertext{2. Boussinesq type systems}
%--------------------------------------------- 
\label{B1}\tag*{\text{$B_1$}} 
&\left\{\begin{aligned}
 u_t &=~~u_{xx}+(u+v)^2,\\
 v_t &= -v_{xx}-(u+v)^2,
\end{aligned}\right.\\
%--------------------------------------------- 
\label{B2}\tag*{\text{$B_2$}}  
&\left\{\begin{aligned}
 u_t &=~~u_{xx}+2 vv_x,\\
 v_t &= -v_{xx}+u_x,
\end{aligned}\right.\\
%--------------------------------------------- 
\label{B3}\tag*{\text{$B_3$}}  
&\left\{\begin{aligned}
 u_t &=~~u_{xx}+6(u+v)v_x-6(u+v)^3,\\
 v_t &= -v_{xx}+6(u+v)u_x+6(u+v)^3,
\end{aligned}\right.\\
%--------------------------------------------- 
\label{B4}\tag*{\text{$B_4$}}  
&\left\{\begin{aligned}
 u_t &=~~u_{xx}+2vv_x,\\
 v_t &= -v_{xx}+2uu_x.
\end{aligned}\right.
\end{align}
\end{theorem}

\begin{remark}
Some of the above systems can be generalized by adding of lower weight terms preserving the integrability. A classification of integrable inhomogeneous polynomial systems can be found, e.g.~in \cite{SW}.
\end{remark}

The goal of the present paper is to find all noncommutative generalizations with conservation laws for the systems from the above lists. Our approach is similar to the construction method of integrable non-Abelian ODE used in \cite{SW2}. We postulate that: 
\begin{itemize}
\item[\bf 1.] a noncommutative generalization is polynomial, homogeneous and admits the scaling group
\[
  (x,\ t, \, {\bf u}, \, {\bf v})\longrightarrow (\tau^{-1}x, \ \tau^{-\mu} t, \, \tau^{\nu_1}  {\bf u},  \tau^{\nu_2}  {\bf v})
\]
with $\mu$, $\nu_1$ and $\nu_2$ which are the same as for the original system. If the weights of the scalar system contain an arbitrary parameter, like in the case of \eqref{NLS}, then we assume that this parameter is preserved also for the noncommutative generalization;

\item[\bf 2.] the generalization turns into the original system under substitution of commuting variables instead of noncommuting ones (or, less formally, for the case of $1\times1$ matrices); 

\item[\bf 3.] for any homogeneous conserved density of the original system, there exists a homogeneous conserved density of its non-Abelian analog, which turns into it under substitution of commuting variables instead of noncommuting ones.  
\end{itemize}

Of course, in practice we make use only of a finite subset of simplest conservation laws of the scalar system. Their number depends on the number of indeterminate parameters in the noncommutative generalization which we are looking for. We stop comparing densities as soon as all parameters are fixed; after that we turn to the search of zero curvature representations for the obtained non-Abelian analogs in order to prove their integrability.

Noncommutative analogs exist for almost all systems from the list, with the exceptions of the system \ref{B1} (see Section \ref{s:B2}) and the family \ref{S5} with generic values of parameters. Moreover, many systems admit more than one generalization. 

\begin{remark}\label{family}
The family $S_5(\alpha,\beta)$ is curious enough and deserves comments. The parameters in this family can be scaled, so that only the ratio $\tau =\alpha:\beta$ is important. It turns out that noncommutative generalizations exist only for $\tau=1/2$, $\tau=0$ and $\tau=\infty$. In the scalar case, the value $\tau$ is changed under a linear transform of the variables $p,q$ related with $u,v$ by the differential substitution
\begin{equation}\label{pquv}
 p_x=u v,\qquad e^q=\frac{u}{v}.
\end{equation}
It is easy to prove that in these variables the system takes the form
\[
 p_t=p_xq_x+(\alpha+\beta)p^2_x,\qquad
 q_t=\frac{p_{xxx}}{p_x}-\frac{p^2_{xx}}{2p^2_x}
     +\frac{q^2_x}{2} -2(\alpha-\beta)p_xq_x -2\alpha(2\alpha-\beta)p^2_x
\]
and that it is invariant under the change \cite[example 8.1]{MikShaYam87} 
\[
 \tilde p=p,\qquad \tilde q=q-2kp,\qquad \tilde\alpha=\alpha+k,\qquad \tilde\beta=\beta+k, \qquad \mbox{where} \quad k\in\mathbb{C}.
\]
Applying the transformation inverse to (\ref{pquv}), we arrive at the system $S_5(\tilde\alpha,\tilde\beta)$ for the variables $\tilde u,\tilde v$, with a new value of $\tau$. The value $\tau=1$ is invariant and corresponds to a linearizable system (cf. Section \ref{linsys}). The symmetries and conservation laws corresponding to different values $\tau\ne 1$ are also related by the above nonlocal transformation. Apparently the values $\tau=1/2$, $0$, $\infty$ may be distinguished from the point of view of some additional structures besides the symmetry approach. It should be noted that, historically, exactly these values appeared as separate scalar systems DNLS-I, DNLS-II and DNLS-III. For consistency, we keep the notation $S_5(\alpha,\beta)$ for their non-Abelian analogs, despite the fact that there are no common family in this case.
\end{remark}
 
A trivial source of multiple non-Abelian generalizations is related with a discrete symmetry group which leaves the scalar system invariant, but changes its noncommutative analog. A most important example is the $\mathbb{C}$-linear involution $\star$ on ${\cal A}$ defined as follows:
\begin{equation}\label{star1}
 u^\star=u\,,\qquad v^\star=v\,,\qquad 
 (a\,b)^\star=b^\star\,a^\star\,,\qquad a,b\in {\cal A}.
\end{equation}
It is clear that this involution applied to any non-Abelian generalization gives some another, possibly different, generalization of the same scalar system. Another possible transformation may be related with a scaling of unknown variables. We consider non-Abelian analogs related by the involution \eqref{star1} or by a scaling as equivalent.

\begin{theorem}\label{th:noncom}
If a non-Abelian analog of one of the scalar systems \ref{S1}--\ref{S5}, \ref{B1}--\ref{B4} satisfies the conditions {\bf 1}--{\bf 3} then it is equivalent to one of the systems listed in the second column of the Table \ref{table:N}, with explicit equations for these systems given in Section \ref*{s:noncom}.
\end{theorem}

\begin{table}[t]
\[
\begin{array}{llcc}
              &                                 & N & \text{weights of }u,v\\
\hline                                                                
\ref{S1}      & \ref{S'1}                       & 1 & (\nu,2-\nu) \\
\ref{S2}      & \ref{S'2}, \ref{S''2}           & 4 & (1,1)       \\
\ref{S3}      & \ref{S'3}                       & 2 & (2,1)       \\
\ref{S4}      & \ref{S'4}, \ref{S''4}           & 4 & (1,1)       \\
\ref{S5}(\alpha,\beta) &                        & 0 & (\nu,1-\nu) \\
\ref{S5}(1,2) & \ref{S'512}, \ref{S''512}       & 3 &             \\
\ref{S5}(0,1) & \ref{S'501}, \ref{S''501}, \ref{S'''501} & 6 &  \\
\ref{S5}(1,0) & \ref{S'510}, \ref{S''510}       & 3 &             \\
\hline         
\ref{B1}      &                                 & 0 & (2,2)       \\
\ref{B2}      & \ref{B'2}                       & 2 & (3,2)       \\
\ref{B3}      & \ref{B'3}, \ref{B''3}           & 4 & (1,1)       \\
\ref{B4}      & \ref{B'4}, \ref{B''4}           & 8 & (1,1)       \\
\hline
\end{array}
\]
\caption{The notations of noncommutative analogs. The last two columns contain the total number $N$ of analogs (that is, without taking the involution and the scaling into account) and the weights of homogeneity.}
\label{table:N}
\end{table}

Generally speaking, the existence of several higher conservation laws does not guarantee integrability in any sense. In order to make sure that the found noncommutative systems are integrable indeed, we find for each system a zero curvature representation
\begin{equation}\label{ZCR}
 U_t = V_x + [V,\,U],
\end{equation}
where $U$ and $V$ are homogeneous polynomial matrices depending on the spectral parameter $\lambda$. These representations are also given in Section \ref{s:noncom}.

Note that in the scalar case, the zero curvature representations admit transformations of the form
\begin{equation}\label{shift}
 U\mapsto U+\rho\, I,\qquad V\mapsto V+\sigma\, I,
\end{equation}
where $\rho_t=\sigma_x$ is any conservation law of the system. However, in the noncommutative situation this transformation is not generally allowed. In this case the elements on the diagonal turn out to be more rigidly fixed and there is only a finite number of choices which bring to different noncommutative analogs.

In Section \ref{s:sym}, we give noncommutative generalizations of linearizable systems of the form \eqref{sys2} and differential substitutions relating them with linear systems. This section contains also examples of master-symmetries and B\"acklund auto-transformations for some of non-Abelian systems presented in Section \ref{s:noncom}. 

\section{The scheme of the proof of the main theorem}
\label{s:proof}

In this section, we outline the proof of the Theorem \ref{th:noncom} by example of the system \ref{B2}. The simplest conserved densities for it are of the form
\begin{gather*}
 \rho_2 = v, \qquad  \rho_3=u, \qquad \rho_5 = u v, \qquad 
 \rho_6= \frac{3}{2} u^2 + v^3 - 3 u v_x,\\
 \rho_8= u u_{xx} + u^2 v + \frac{1}{3}v^4 - 2 u v v_x + v^2 v_{xx} - 2 u v_{xxx}.
\end{gather*}
Here the subscripts denote the weights of the densities (there are no densities of weights of the form $3n+1$ for this system). The most general nocommutative system which is homogeneous with respect to the same weights as for \ref{B2} and which turns into \ref{B2} in the commutative case is of the form
\begin{align}\label{nabB2}
\left\{\begin{aligned}
 {\bf u}_t &= {\bf u}_{xx}+\alpha {\bf v}{\bf v}_x + (2-\alpha) {\bf v}_x {\bf v} + \beta ({\bf u} {\bf v}-{\bf v} {\bf u}),\\
 {\bf v}_t &= -{\bf v}_{xx}+{\bf u}_x.
\end{aligned}\right.
\end{align}
The parameters $\alpha$ and $\beta$ must be determined by condition that \eqref{nabB2} admits non-Abelian conserved densities which turn into the above $\rho_i$ in the scalar case.

The most general non-Abelian density corresponding to $\rho_2$ is of the form $\bar\rho_2= {\bf v}$. It is clear that  
\[
 D_t(\bar \rho_2)= -{\bf v}_{xx}+{\bf u}_x \in\operatorname{Im}D \subset {\cal T},
\]
that is, this gives no information about the parameters $\alpha$, $\beta$. The conditions that $\bar\rho_3 = {\bf u}$ and $\bar\rho_5={\bf u \, v}$ are non-Abelian densities are satisfied identically as well. For instance, for $\bar\rho_3$ we have
\[
 D_t(\bar\rho_3)= D({\bf u}_x+{\bf v}^2)+(\alpha-1)[{\bf v},{\bf v}_x] 
  +\beta [{\bf u},{\bf v}] \in {\cal T}.
\]
Let us take the next density. One can easily verify that any noncommutative homogeneous polynomial of weight 6 which coincide with $\rho_6$ in the scalar case is of the form
\[
 \frac{3}{2} {\bf u}^2 + k_1 ({\bf u}_x {\bf v} - {\bf v} {\bf u}_x) 
  +k_2 ({\bf u} {\bf v}_x - {\bf v}_x {\bf u}) - 3 {\bf v}_x {\bf u} 
  +k_3 ({\bf v} {\bf v}_{xx} - {\bf v}_{xx} {\bf v}) + {\bf v}^3, \qquad k_i\in\mathbb{C}.
\]
Since non-Abelian densities are defined modulo ${\cal T}$, we may set $k_1=k_2=k_3=0$ and assume without loss of generality that 
\[
 \bar\rho_6 = \frac{3}{2} {\bf u}^2 + {\bf v}^3 - 3 {\bf v}_x {\bf u}.
\]
It is not difficult to prove that
\[
 D_t(\bar\rho_6)= D(K_0) + \alpha K_1+\beta K_2 + K_3 
  + 3 (\alpha-\beta-1)({\bf v}_x {\bf u} {\bf v} -{\bf v}_x {\bf v} {\bf u}), 
\]
where  
\begin{gather*}
 K_0= -D({\bf v}^3) + 3 {\bf v}_{xx} {\bf u} - 3 {\bf v}_x {\bf u}_x 
      +{\bf u} {\bf v}^2+{\bf v}{\bf u}{\bf v}+{\bf v}^2{\bf u},\\
 K_1= 3 [{\bf v}_x, {\bf v}_x {\bf v}] - \frac{3}{2} [{\bf v}_x, {\bf u} {\bf v}]
     +\frac{3}{2} [{\bf v}, {\bf v}_x {\bf u}]
     -\frac{3}{2} [{\bf u}, {\bf v}_x {\bf v}],\qquad 
 K_2= \frac{3}{2} [{\bf u}^2, {\bf v}],\\
 K_3= \frac{3}{2} [{\bf u}, {\bf u}_{xx} ]+2 [{\bf v}, {\bf v}_x^2] 
      - 2 [{\bf v}_x, {\bf v}_x{\bf v}]
      +[{\bf u},{\bf v}_x{\bf v}]+[{\bf v}_x,{\bf u}{\bf v}]
      -[{\bf v},{\bf u}{\bf v}_x]-[{\bf v},{\bf v}_x{\bf u}].
\end{gather*}
Let us prove that the polynomial $P={\bf v}_x {\bf u} {\bf v} -{\bf v}_x {\bf v} {\bf u}$ does not belong to ${\cal T}$. Since all terms in $D_t(\bar \rho_3)$ except the last one belong to ${\cal T}$, this will imply that
\[
 \alpha=\beta+1.
\] 
In order to prove this we use the following property of the non-Abelian variational derivatives:
\[
 \frac{\delta}{\delta {\bf u}}(a)=\frac{\delta}{\delta {\bf v}}(a) = 0, \quad \forall a\in {\cal T}.
\] 
The variational principle for functionals of the form \eqref{funk} implies the following definition of the variational derivatives. One has to perform the substitution ${\bf u}_i\mapsto {\bf u}_i +\varepsilon \Delta^u_i$, ${\bf v}_i\mapsto {\bf v}_i + \varepsilon \Delta^v_i$ into the arguments of the polynomial $a$ and to compute the coefficient at $\varepsilon$ (that is, the differential $\Delta a$). For the polynomial $P$, we obtain
\[
 \Delta P = \Delta^v_x {\bf u}{\bf v}+{\bf v}_x \Delta^u {\bf v}+{\bf v}_x {\bf u} \Delta^v
  -\Delta^v_x {\bf v}{\bf u}-{\bf v}_x \Delta^v {\bf u}-{\bf v}_x {\bf v} \Delta^u.
\]
Next, one has to bring $\Delta a$ to the form $\Delta^u R + \Delta^v S$ by subtracting commutators and total $x$-derivatives. The polynomials  $R$ and $S\in {\cal A}$ are uniquely defined and are denoted $\frac{\delta a}{\delta {\bf u}}$ and $\frac{\delta a}{\delta {\bf v}}$, respectively. In our example, we have 
\begin{gather*}
 \Delta P - D(\Delta^v {\bf u}{\bf v} ) - [{\bf v}_x, \Delta^u {\bf v}] 
  - [{\bf v}_x {\bf u}, \Delta^v]+D(\Delta^v {\bf v}{\bf u} ) 
  + [{\bf v}_x, \Delta^v {\bf u}] + [{\bf v}_x {\bf v}, \Delta^u] = \\
 = \Delta^u [{\bf v}, {\bf v}_x] 
  + \Delta^v \bigl( [{\bf v}, {\bf u}_x]+2 [{\bf v}_x, {\bf u}] \bigr).
\end{gather*}
Therefore,
\[
 \frac{\delta}{\delta {\bf u}}(P) = [{\bf v}, {\bf v}_x], \qquad 
 \frac{\delta}{\delta {\bf v}}(P) = [{\bf v}, {\bf u}_x] +2[{\bf v}_x, {\bf u}].  
\]
Since the variational derivatives do not vanish, $P$ does not belong to ${\cal T}$.

\begin{remark} 
The result of the above computation procedure for the variational derivatives can be given by an explicit expression. Let $P$ be an arbitrary polynomial of noncommuting variables ${\bf u}^1_i,\dots,{\bf u}^N_i$ (where subscripts denote the order of derivatives with respect to $x$). Let us denote, for any monomial $p={\bf u}^{s_1}_{i_1}\dots{\bf u}^{s_m}_{i_m}$, 
\[
 L_k(p)={\bf u}^{s_1}_{i_1}\dots{\bf u}^{s_{k-1}}_{i_{k-1}},\qquad C_k(p)={\bf u}^{s_k}_{i_k},\qquad
 R_k(p)={\bf u}^{s_{k+1}}_{i_{k+1}}\dots{\bf u}^{s_m}_{i_m},
\]
so that always $p=L_k(p)C_k(p)R_k(p)$. Then, given a polynomial $P=\sum_j\alpha_jp_j$, we have
\[
 \frac{\delta P}{\delta{\bf u}^s}= \sum_j\alpha_j \sum_{k,i:\,C_k(p_j)={\bf u}^s_i}(-D)^i(R_k(p_j)L_k(p_j)),\qquad s=1,\dots,N. 
\]
\end{remark}

The parameter $\beta$ is determined at the next step. It is not difficult to prove that the most general ansatz for the density $\bar\rho_8$ can be taken in the form
\[
 \bar\rho_8 = {\bf u}_{xx} {\bf u} - 2 {\bf v}_{xxx} {\bf u}+ {\bf v} {\bf u}^2 
  +\zeta{\bf v}_x{\bf u}{\bf v} - (2+\zeta){\bf v}_x{\bf v}{\bf u}
  +{\bf v}_{xx}{\bf v}^2+\frac{1}{3} {\bf v}^4.
\]
Notice that a free parameter $\zeta$ appeared in this density for the first time. The presence of such parameters leads to nonlinear algebraic relations for the coefficients of the noncommutative generalization that we are looking for. A direct computation shows that vanishing of the variational derivatives of the expression $D_t(\bar\rho_8)$ is equivalent to the system of equations
\[
 \zeta+\beta+1=0, \qquad 3 \beta(1+\zeta) = 1.
\]
From here we obtain $\beta=\pm i\sqrt{3}/3$. Two non-Abelian systems corresponding to different signs of $\beta$ are related by involution \eqref{star1}.

As the weights of $u$ and $v$ decrease, the number of monomials participating in the right-hand sides of the noncommutative system and in expressions for their densities increases and the calculations become more cumbersome. However, they remain fairly straightforward and, what is important, a large part of the relations for the unknown coefficients are linear equations. As a rule, quadratic relations arise only at the last step and are easily analyzed.

\section{Non-Abelian systems and their zero curvature representations}
\label{s:noncom}  

In this section we present all non-Abelian analogs admitting conserved densities for systems from the lists $S_n$, $B_n$. In order to justify their integrability we provide the zero curvature representations \eqref{ZCR}. The search of the matrices $U$ and $V$ was performed under assumption that their entries are homogeneous polynomials from ${\cal A}$. Another assumption was that, as in the scalar case, the size of the matrices is $2\times 2$ for the NLS type systems and $3\times 3$ for the Boussinesq type systems. This method amounts to solving a nonlinear system of algebraic equations for the indeterminate coefficients of the elements of $U$ and $V$. It turns out that this system is solvable for all systems under considerations, although bringing the answer to an `elegant' form may require some efforts.

In what follows, ${\bf I}$ denotes the unity of the algebra ${\cal A}$ and $I$ denotes the $2\times 2$ or $3\times 3$ matrix with ${\bf I}$ on its diagonal.

\subsection{Nonlinear Schr\"odinger equation \texorpdfstring{$S_1$}{S1}}

The system \ref{S1} admits one noncommutative analog:
\begin{align}
\label{S'1}\tag*{\text{$S'_1$}}
&\left\{\begin{aligned}
 {\bf u}_t &=~~{\bf u}_{xx}+2{\bf u}{\bf v}{\bf u},\\
 {\bf v}_t &= -{\bf v}_{xx}-2{\bf v}{\bf u}{\bf v}.
\end{aligned}\right.
\end{align}
It is clear that this system is invariant with respect to the involution \eqref{star1}. The matrices of the zero curvature representation are of the form (see e.g. \cite{sv15})
\[ 
 U=\begin{pmatrix}
   \lambda{\bf I} & -{\bf v} \\
   {\bf u} & -\lambda{\bf I}
  \end{pmatrix},\qquad
 V=-2\lambda U
  +\begin{pmatrix}
   -{\bf v}{\bf u} & {\bf v}_x\\
   {\bf u}_x & {\bf u}{\bf v}
  \end{pmatrix}. 
\]

\subsection{Levi system \texorpdfstring{$S_2$}{S2}}

The system \ref{S2} \cite{Levi_1981} admits two very similar, yet non-equivalent noncommutative generalizations (and two more are obtained from
them by involution):
\begin{align}
\label{S'2}\tag*{\text{$S'_2$}}
&\left\{\begin{aligned}
 {\bf u}_t &=~~{\bf u}_{xx}+2{\bf u}_x{\bf u}+2({\bf vu})_x+2[{\bf vu},{\bf u}],\\
 {\bf v}_t &= -{\bf v}_{xx}+2{\bf v}_x{\bf v}+2({\bf uv})_x+2[{\bf v},{\bf uv}],
\end{aligned}\right.\\[1.5mm]
\label{S''2}\tag*{\text{$S''_2$}}
&\left\{\begin{aligned}
 {\bf u}_t &=~~{\bf u}_{xx}+2{\bf uu}_x+2({\bf vu})_x+2[{\bf u},{\bf vu}],\\
 {\bf v}_t &= -{\bf v}_{xx}+2{\bf v}_x{\bf v}+2({\bf vu})_x+2[{\bf v},{\bf vu}].
\end{aligned}\right.
\end{align}
The difference between these two systems can be observed under the reduction ${\bf v}=-{\bf u}$ in their third order symetries, which leads to two different generalizations of the mKdV equation: the symmetry of \ref{S'2} turns into \eqref{mat2} and the symmetry of \ref{S''2} turns into \eqref{matmkdv}.

The matrices ot the zero curvature representation for \ref{S'2} are
\[ 
 U=\begin{pmatrix}
   0 & {\bf I}\\
   \lambda{\bf v} & \lambda{\bf I}+{\bf u}-{\bf v}
  \end{pmatrix},\qquad 
 V=(2{\bf u}I-U)U
   +\begin{pmatrix}
    0 & 0\\
   -\lambda{\bf v}_x& {\bf u}_x+{\bf v}_x
   \end{pmatrix},
\]
the matrices for the systems \ref{S''2} are
\[ 
 U=\begin{pmatrix}
   -{\bf u} & {\bf I}\\
  \lambda{\bf v} & \lambda{\bf I}-{\bf v}
  \end{pmatrix},\qquad 
 V=-U^2-\begin{pmatrix}
   {\bf u}_x+2{\bf vu} & 0\\
   \lambda{\bf v}_x & -{\bf v}_x+2{\bf vu}
   \end{pmatrix}. 
\]

\subsection{Kaup--Broer system \texorpdfstring{$S_3$}{S3}}

The system \ref{S3} has, up to the involution, one noncommutative analog (see e.g. \cite{kuper}):
\begin{align}\label{S'3}\tag*{\text{$S'_3$}}
\left\{\begin{aligned}
 {\bf u}_t &=~~{\bf u}_{xx}+2({\bf vu})_x,\\
 {\bf v}_t &= -{\bf v}_{xx}+2{\bf v}_x{\bf v}+2{\bf u}_x+2[{\bf v},{\bf u}].
\end{aligned}\right.
\end{align}
The auxiliary linear problems looks exactly the same as in the scalar case:
\[
 \psi_{xx}+({\bf v}-\lambda)\psi_x+{\bf u}\psi=0,\qquad \psi_t=\psi_{xx}+2{\bf v}\psi_x.
\]
The consistency condition for these equations is equivalent to \ref{S'3}. This can be cast into the form of representation (\ref{ZCR}) with the matrices
\[
 U=\begin{pmatrix}  
   0 & {\bf I} \\
  -{\bf u} & \lambda{\bf I}-{\bf v}
 \end{pmatrix},\qquad
 V=\begin{pmatrix}
  -{\bf u} & \lambda{\bf I}+{\bf v} \\
  -\lambda{\bf u}-{\bf u}_x-{\bf vu} & \lambda^2{\bf I}+{\bf v}_x-{\bf u}-{\bf v}^2
 \end{pmatrix}.
\]

\subsection{System \texorpdfstring{$S_4$}{S4}}

Up to the involution, the system \ref{S4} admits two noncommutative analogs:
\begin{align}
\label{S'4}\tag*{\text{$S'_4$}}
&\left\{\begin{aligned}
 {\bf u}_t &=~~{\bf u}_{xx}+2{\bf u}_x({\bf u}+{\bf v}),\\
 {\bf v}_t &= -{\bf v}_{xx}+2({\bf u}+{\bf v}){\bf v}_x,
\end{aligned}\right.\\[1.5mm]
\label{S''4}\tag*{\text{$S''_4$}}
&\left\{\begin{aligned}
 {\bf u}_t &=~~{\bf u}_{xx}+2({\bf u}+{\bf v}){\bf u}_x+2[{\bf v}_x,{\bf u}]
             -2{\bf u}^2{\bf v}+4{\bf u}{\bf v}{\bf u}-2{\bf v}{\bf u}^2,\\
 {\bf v}_t &= -{\bf v}_{xx}+2{\bf v}_x({\bf u}+{\bf v})+2[{\bf v},{\bf u}_x]
             +2{\bf u}{\bf v}^2-4{\bf v}{\bf u}{\bf v}+2{\bf v}^2{\bf u}.
\end{aligned}\right.
\end{align}
The zero curvature representation matrices for \ref{S'4} are of the form
\begin{align*}
 & U=\begin{pmatrix}
   {\bf u} & (\lambda{\bf I}-{\bf u})(\lambda{\bf I}-{\bf v})\\
   {\bf I} & {\bf v}
 \end{pmatrix},\\
 & V=U^2+2\lambda U-\lambda^2I
  +\begin{pmatrix}
   {\bf u}_x & \lambda({\bf v}_x-{\bf u}_x)+{\bf u}_x{\bf v}-{\bf uv}_x\\
   0 & -{\bf v}_x
  \end{pmatrix};
\end{align*}
the matrices for \ref{S''4} read
\begin{align*}
 & U=\begin{pmatrix}
   -{\bf v} & (\lambda{\bf I}-{\bf u})(\lambda{\bf I}-{\bf v})\\
   {\bf I} & -{\bf u}
 \end{pmatrix},\\[1.5mm]
 & V=-U^2+2\lambda U+(\lambda^2{\bf I}+2[{\bf u},{\bf v}])I
  +\begin{pmatrix}
   {\bf v}_x & \lambda({\bf v}-{\bf u})_x +{\bf u}_x{\bf v}-{\bf uv}_x\\
   0 & -{\bf u}_x
  \end{pmatrix}.
\end{align*}

\subsection{DNLS-I, system \texorpdfstring{$S_5(1,2)$}{S5(1,2)}}

The system \ref{S5}$(1,2)$ (DNLS-I or the Kaup--Newell system \cite{Kaup_Newell_1978}) admits two noncommutative generalizations (the first one is symmetric with respect to the involution and the second one is not):
\begin{align}
\label{S'512}\tag*{\text{$S'_5(1,2)$}}
&\left\{\begin{aligned}
 {\bf u}_t &=~~{\bf u}_{xx}+2({\bf u}{\bf v}{\bf u})_x,\\
 {\bf v}_t &= -{\bf v}_{xx}+2({\bf v}{\bf u}{\bf v})_x,
\end{aligned}\right.\\[1.5mm]
\label{S''512}\tag*{\text{$S''_5(1,2)$}}
&\left\{\begin{aligned}
 {\bf u}_t &=~~{\bf u}_{xx}+2{\bf u}^2{\bf v}_x+2{\bf u}_x{\bf u}{\bf v}
     +2{\bf u}_x{\bf v}{\bf u}+2[{\bf u}^2{\bf v}^2,{\bf u}],\\
 {\bf v}_t &= -{\bf v}_{xx}+2{\bf u}_x{\bf v}^2+2{\bf u}{\bf v}{\bf v}_x
     +2{\bf v}{\bf u}{\bf v}_x -2[{\bf v},{\bf u}^2{\bf v}^2].
\end{aligned}\right.
\end{align}
The matrices of zero curvature representation for \ref{S'512}: 
\[
 U= 2\lambda\begin{pmatrix}  
   \lambda{\bf I} & {\bf u} \\
   -{\bf v} & -\lambda{\bf I}
 \end{pmatrix},\quad
 V= 4\lambda^2U+2\lambda\begin{pmatrix}
  2\lambda{\bf uv} & {\bf u}_x+2{\bf uvu} \\
  {\bf v}_x-2{\bf vuv} & -2\lambda{\bf vu}
 \end{pmatrix}; 
\] 
for \ref{S''512}: 
\begin{align*}
 &U= \begin{pmatrix}  
   2\lambda^2{\bf I}+{\bf u}{\bf v} & 2\lambda{\bf u} \\
   -2\lambda{\bf v} & -2\lambda^2{\bf I}+{\bf u}{\bf v}
 \end{pmatrix},\\
 &V= U^2 +(4\lambda^2{\bf uv}+{\bf u}_x{\bf v}-{\bf uv}_x+2{\bf u}^2{\bf v}^2)I
  +2\lambda\begin{pmatrix}
   2\lambda^3{\bf I} & 4\lambda^2{\bf u}+{\bf u}_x \\
   -4\lambda^2{\bf v}+{\bf v}_x & -6\lambda^3{\bf I}
 \end{pmatrix}. 
\end{align*}

\subsection{DNLS-II, system \texorpdfstring{$S_5(0,1)$}{S5(0,1)}}

The system \ref{S5}$(0,1)$ is known as DNLS-II or the Chen--Lee--Liu system \cite{Chen_Lee_Liu_1979}. It has there noncommutative analogs (and three more are obtained by involution):
\begin{align}
\label{S'501}\tag*{\text{$S'_5(0,1)$}}
&\left\{\begin{aligned}
 {\bf u}_t &=~~{\bf u}_{xx}+2{\bf uvu}_x,\\
 {\bf v}_t &= -{\bf v}_{xx}+2{\bf v}_x{\bf uv},
\end{aligned}\right.\\[1.5mm]
\label{S''501}\tag*{\text{$S''_5(0,1)$}}
&\left\{\begin{aligned}
 {\bf u}_t &=~~{\bf u}_{xx}+2{\bf v}{\bf u}{\bf u}_x
        +2[{\bf v}_x{\bf u}-{\bf v}[{\bf v},{\bf u}]{\bf u},{\bf u}],\\
 {\bf v}_t &= -{\bf v}_{xx}+2{\bf v}_x{\bf v}{\bf u}
        +2[{\bf v},{\bf v}{\bf u}_x+{\bf v}[{\bf v},{\bf u}]{\bf u}],
\end{aligned}\right.\\[1.5mm]
\label{S'''501}\tag*{\text{$S'''_5(0,1)$}}
&\left\{\begin{aligned}
 {\bf u}_t &=~~{\bf u}_{xx}+2{\bf u}_x{\bf vu}
      +2[{\bf uv},{\bf u}_x]+2[{\bf uv}_x,{\bf u}]
      -2{\bf u}({\bf u}^2{\bf v}-2{\bf u}{\bf v}{\bf u}+{\bf v}{\bf u}^2){\bf v},\\
 {\bf v}_t &= -{\bf v}_{xx}+2{\bf vuv}_x
      +2[{\bf v}_x,{\bf uv}]+2[{\bf v},{\bf u}_x{\bf v}]
      +2{\bf u}({\bf u}{\bf v}^2-2{\bf v}{\bf u}{\bf v}+{\bf v}^2{\bf u}){\bf v}.
\end{aligned}\right.
\end{align}
The zero curvature representations for these systems are given by the following matrices. For \ref{S'501}: 
\begin{align*}
 & U=\begin{pmatrix}
   -\lambda^2{\bf I} & \lambda{\bf u}\\
   \lambda{\bf v} & -{\bf vu}
 \end{pmatrix},\quad 
 V=-U^2-2\lambda^2U
  +\begin{pmatrix}
   0 & \lambda{\bf u}_x\\
   -\lambda{\bf v}_x & {\bf v}_x{\bf u}-{\bf vu}_x
   \end{pmatrix};
\end{align*}
for \ref{S''501}: 
\begin{align*}
 & U=\begin{pmatrix}
   -\lambda^2{\bf I}+[{\bf u},{\bf v}] & \lambda{\bf u}\\
   \lambda{\bf v} & -{\bf vu}
 \end{pmatrix},\\ 
 & V= -U^2 -2\lambda^2U +2{\bf v}[{\bf u},{\bf v}]{\bf u}I
  +\begin{pmatrix}
   [{\bf u}_x,{\bf v}]-[{\bf u},{\bf v}_x] & \lambda{\bf u}_x\\
   -\lambda{\bf v}_x & {\bf v}_x{\bf u}-{\bf vu}_x
   \end{pmatrix};
\end{align*}
and for \ref{S'''501}: 
\begin{align*}
 & U=\begin{pmatrix}
   -\lambda^2{\bf I} & \lambda{\bf u}\\
   \lambda{\bf v} & -{\bf u}{\bf v}
 \end{pmatrix},\\ 
 & V= -U^2 -2\lambda^2U 
  +\begin{pmatrix}
    0 & \lambda{\bf u}_x+2\lambda[{\bf uv},{\bf u}]\\
   -\lambda{\bf v}_x-2\lambda[{\bf uv},{\bf v}] & 
   -2\lambda^2[{\bf u},{\bf v}]+{\bf uv}_x-{\bf u}_x{\bf v}+2{\bf u}[{\bf u},{\bf v}]{\bf v}
   \end{pmatrix}.
\end{align*}
It is easy to see that in the scalar case all three representations are equivalent up to the diagonal shift \eqref{shift} corresponding to the conservation law with the density $uv$.

\subsection{DNLS-III, system \texorpdfstring{$S_5(1,0)$}{S5(1,0)}}

The system \ref{S5}$(1,0)$ is known as DNLS-III or the Gerdjikov--Ivanov system \cite{Gerdjikov_Ivanov_1983}. It has two noncommutative generalizations:
\begin{align}
\label{S'510}\tag*{\text{$S'_5(1,0)$}}
&\left\{\begin{aligned}
 {\bf u}_t &=~~{\bf u}_{xx}+2{\bf uv}_x{\bf u}-2{\bf uvuvu},\\
 {\bf v}_t &= -{\bf v}_{xx}+2{\bf vu}_x{\bf v}+2{\bf vuvuv},
\end{aligned}\right.\\[1.5mm]
\label{S''510}\tag*{\text{$S''_5(1,0)$}}
&\left\{\begin{aligned}
 {\bf u}_t &=~~{\bf u}_{xx}+2{\bf u}^2{\bf v}_x-2{\bf u}^3{\bf v}^2
             +2[{\bf u}_x+{\bf u}^2{\bf v},{\bf uv}],\\
 {\bf v}_t &= -{\bf v}_{xx}+2{\bf u}_x{\bf v}^2+2{\bf u}^2{\bf v}^3
             -2[{\bf v}_x-{\bf u}{\bf v}^2,{\bf uv}].
\end{aligned}\right.
\end{align}
The zero curvature representation for \ref{S'510} is defined by the matrices
\begin{align*}
 & U= \begin{pmatrix}
   -{\bf uv} & \lambda{\bf u}\\
   -\lambda{\bf v} & \lambda^2{\bf I}+{\bf vu}
 \end{pmatrix},\\[1.5mm]
 & V= -\lambda^2U+
  \begin{pmatrix}
   {\bf uv}_x-{\bf u}_x{\bf v}-({\bf uv})^2 & \lambda{\bf u}_x\\
   \lambda{\bf v}_x & {\bf vu}_x-{\bf v}_x{\bf u}+({\bf vu})^2
   \end{pmatrix};
\end{align*}
similarly, for \ref{S''510} we have
\begin{align*}
 & U= \begin{pmatrix}
    0 & \lambda{\bf u}\\
   -\lambda{\bf v} & \lambda^2{\bf I} +{\bf uv}+{\bf v}{\bf u}
 \end{pmatrix},\\[1.5mm]
 & V= -\lambda^2U+\begin{pmatrix}
   \lambda^2{\bf uv} & \lambda{\bf u}_x+\lambda[{\bf u},{\bf uv}]\\
   \lambda{\bf v}_x +\lambda[{\bf v},{\bf uv}]& 
   \lambda^2{\bf uv}
     +{\bf u}_x{\bf v}-{\bf uv}_x +{\bf vu}_x-{\bf v}_x{\bf u} 
     +2{\bf u}^2{\bf v}^2-[{\bf u},{\bf v}]^2
   \end{pmatrix}.
\end{align*}

\subsection{The system \texorpdfstring{$B_2$}{B2}}
\label{s:B2}

The Boussinesq type system \ref{B2} has one noncommutative analog, up to the involution (a detailed analysis of this example is in Section \ref{s:proof}):
\begin{equation}\label{B'2}\tag*{\text{$B'_2$}} 
\left\{\begin{aligned}
 {\bf u}_t &= {\bf u}_{xx}+{\bf v}_x{\bf v}+{\bf vv}_x
     +i\frac{\sqrt{3}}{3}[{\bf u}-{\bf v}_x,{\bf v}],\\
 {\bf v}_t &= -{\bf v}_{xx}+{\bf u}_x.
\end{aligned}\right.
\end{equation}

\begin{remark}
Although there are no local non-Abelian analogs for \ref{B1}, this system admits the nonlocal generalization
\begin{equation}\label{B'1} 
\left\{\begin{aligned}
 {\bf u}_t &=~~{\bf u}_{xx}+({\bf u}+{\bf v})^2+{\bf w},\\
 {\bf v}_t &= -{\bf v}_{xx}-({\bf u}+{\bf v})^2-{\bf w},
\end{aligned}\right.\qquad
 {\bf w}_x = i\frac{\sqrt{3}}{3}[{\bf u}_x-{\bf v}_x,\,{\bf u}+{\bf v}],
\end{equation}
which is related with \ref{B'2} by non-invertible differential substitution ${\bf u}=4\tilde{\bf u}_x$, ${\bf v}=2\tilde{\bf u}+2\tilde{\bf v}$ (variables with tilde correspond to (\ref{B'1})).  A systematic description of differential substitutions connecting the non-Abelian systems presented in the paper is a separate interesting problem.
\end{remark}

The zero curvature representations for all Boussinesq type systems are more conveniently written in some another variables related with ${\bf u}$ and ${\bf v}$ by some invertible linear changes. Such changes lead out of the class of systems (\ref{sys2}), because they do not preserve the {\em separant} of the system (the matrix at the second order derivatives on the right-hand side). For the system \ref{B'2}, the transformation
\begin{equation}\label{pq}
 \partial_t=-i\sqrt{3}\,\partial_T,\qquad 
 {\bf p}=\frac{i\sqrt{3}}{12}{\bf u}+\frac{3-i\sqrt{3}}{12}{\bf v}_x,\qquad 
 {\bf q}=\frac{1}{6}{\bf v} 
\end{equation}
brings to the system
\begin{equation}\label{B2.pq}
 {\bf p}_T={\bf p}_{xx}-2{\bf q}_{xxx}-6{\bf q}{\bf q}_x+2[{\bf q},{\bf p}],\qquad 
 {\bf q}_T=-{\bf q}_{xx}+\frac{2}{3}{\bf p}_x,
\end{equation}
which serves as the compatibility condition for the third order spectral problem \cite{Kaup_1980}
\[
 \psi_{xxx}+3{\bf q}\psi_x+{\bf p}\psi=\lambda\psi,\qquad \psi_T=\psi_{xx}+2{\bf q}\psi. 
\]
The representation (\ref{ZCR}) for \eqref{B2.pq} is given by the matrices 
\[
 U=\begin{pmatrix}
  0 & {\bf I} & 0\\
  0 & 0 & {\bf I}\\
  \lambda {\bf I}-{\bf p} & -3{\bf q} & 0
 \end{pmatrix},\qquad
 V=\begin{pmatrix}
  2{\bf q} & 0 & {\bf I}\\
  \lambda{\bf I}+2{\bf q}_x-{\bf p} & -{\bf q} & 0\\
  2{\bf q}_{xx}-{\bf p}_x & \lambda{\bf I}+{\bf q}_x-{\bf p} & -{\bf q}
 \end{pmatrix}.
\]
In order to obtain a representation for \ref{B'2}, one has just to replace ${\bf p}$ and ${\bf q}$ according to (\ref{pq}).

One can see that the appearance of $\sqrt{3}$ in \ref{B'2} is not related with the passage to the noncommuting variables, rather this is a price we pay for bringing the system \eqref{B2.pq} to the canonical form (the third derivative is eliminated and the matrix at second derivatives is made diagonal). In fact, the radicals appear already in the scalar case \ref{B2}, hidden in the zero curvature representation, but they cancel in the system itself. In the noncommutative version no cancellation occurs and the radicals become visible.

\subsection{System \texorpdfstring{$B_3$}{B3}}

For \ref{B3}, there are two noncommutative analogs, up to the involution (which amounts to the complex conjugation, like in the previous example): 
\begin{align}
\label{B'3}\tag*{\text{$B'_3$}}
&\left\{\begin{aligned}
 {\bf u}_t &=~~{\bf u}_{xx}+3({\bf u}+{\bf v}){\bf v}_x+3{\bf v}_x({\bf u}+{\bf v})
             -6({\bf u}+{\bf v})^3+i\sqrt{3}[{\bf u}+{\bf v},{\bf v}_x],\\
 {\bf v}_t &= -{\bf v}_{xx}+3({\bf u}+{\bf v}){\bf u}_x+3{\bf u}_x({\bf u}+{\bf v})
             +6({\bf u}+{\bf v})^3-i\sqrt{3}[{\bf u}+{\bf v},{\bf u}_x],
\end{aligned}\right.\\[2mm]
\label{B''3}\tag*{\text{$B''_3$}} 
&\left\{\begin{aligned}
 {\bf u}_t &=~~{\bf u}_{xx}+3({\bf u}+{\bf v}){\bf v}_x+3{\bf v}_x({\bf u}+{\bf v})\\
     &\qquad -6({\bf u}^3+3{\bf uvu}+{\bf uv}^2+{\bf vuv}+{\bf v}^2{\bf u}+{\bf v}^3)\\
     &\qquad +i\sqrt{3}\bigl([{\bf v}_x,{\bf u}-{\bf v}]
             +2[{\bf u}+{\bf v},{\bf u}_x]+6[{\bf u}^2,{\bf v}]+6[{\bf u},{\bf v}^2]\bigr),\\[1.5mm]
 {\bf v}_t &= -{\bf v}_{xx}+3({\bf u}+{\bf v}){\bf u}_x+3{\bf u}_x({\bf u}+{\bf v})\\
     &\qquad +6({\bf u}^3+{\bf u}^2{\bf v}+{\bf uvu}+{\bf vu}^2+3{\bf vuv}+{\bf v}^3)\\
     &\qquad +i\sqrt{3}\bigl([{\bf u}_x,{\bf u}-{\bf v}]
             -2[{\bf u}+{\bf v},{\bf v}_x]-6[{\bf u}^2,{\bf v}]-6[{\bf u},{\bf v}^2]\bigr).
\end{aligned}\right.
\end{align}
The system \ref{B'3} admits the representation (\ref{ZCR}) with
\begin{align*}
 & U=\begin{pmatrix}
  i\sqrt{3}({\bf u}+{\bf v}) &  {\bf I} & 0 \\
   0  & i\sqrt{3}(\varepsilon^2{\bf u}+\varepsilon{\bf v}) & {\bf I} \\
  \lambda{\bf I} & 3i\sqrt{3}(\varepsilon^2{\bf vu}-\varepsilon{\bf uv}) 
     & i\sqrt{3}(\varepsilon{\bf u}+\varepsilon^2{\bf v})
 \end{pmatrix},\\
 & V=-i\sqrt{3}U^2
  +i\sqrt{3}\begin{pmatrix}
  {\bf u}_x-{\bf v}_x & 0 & 0 \\
   0 & \varepsilon^2{\bf u}_x-\varepsilon{\bf v}_x & 0 \\
   0 & {\bf a} & \varepsilon{\bf u}_x-\varepsilon^2{\bf v}_x 
 \end{pmatrix},
\end{align*}
where we denote $\varepsilon=-1/2+i\sqrt{3}/2$ and
\[
 {\bf a}=3\varepsilon({\bf u}{\bf v}_x-{\bf u}_x{\bf v})
 +3\varepsilon^2({\bf v}{\bf u}_x-{\bf v}_x{\bf u})-6i\sqrt{3}({\bf u}+{\bf v})^3.
\]
The matrices for \ref{B''3} are of the form
\begin{align*}
 & \widetilde U=U-i\sqrt{3}({\bf u}+{\bf v})I,\\
 & \widetilde V=V-i\sqrt{3}({\bf u}_x-{\bf v}_x+3({\bf u}+{\bf v})^2)I
  -9\begin{pmatrix}
    0 & 0 & 0 \\
    0 & [{\bf u},{\bf v}] & 0 \\
    0 & {\bf b} & [{\bf u},{\bf v}] 
  \end{pmatrix},
\end{align*}
where
\[
 {\bf b}= {\bf uvu}+\varepsilon{\bf u}^2{\bf v}+\varepsilon^2{\bf vu}^2
         +{\bf vuv}+\varepsilon{\bf uv}^2+\varepsilon^2{\bf v}^2{\bf u}.
\]

\subsection{System \texorpdfstring{$B_4$}{B4}} 

The system \ref{B4} admits two non-equivalent generalizations:
\begin{align}
\label{B'4}\tag*{\text{$B'_4$}} 
&\left\{\begin{aligned}
 {\bf u}_t &=~~{\bf u}_{xx}+{\bf v}{\bf v}_x+{\bf v}_x{\bf v}
             +i\frac{\sqrt{3}}{3}[{\bf v},{\bf u}^2-{\bf v}_x]
             -\frac{1}{3}({\bf u}^2{\bf v}-2{\bf u}{\bf v}{\bf u}+{\bf v}{\bf u}^2),\\
 {\bf v}_t &= -{\bf v}_{xx}+{\bf u}{\bf u}_x+{\bf u}_x{\bf u}
             +i\frac{\sqrt{3}}{3}[{\bf u},{\bf v}^2+{\bf u}_x]
             +\frac{1}{3}({\bf u}{\bf v}^2-2{\bf v}{\bf u}{\bf v}+{\bf v}^2{\bf u}),
\end{aligned}\right.\\[1.5mm]
\label{B''4}\tag*{\text{$B''_4$}} 
&\left\{\begin{aligned}
 {\bf u}_t &=~~{\bf u}_{xx}+{\bf v}{\bf v}_x+{\bf v}_x{\bf v} 
             +{\bf u}^2{\bf v}-2{\bf u}{\bf v}{\bf u}+{\bf v}{\bf u}^2 \\
     &\qquad -i\frac{\sqrt{3}}{3}\bigl(2[{\bf u}+{\bf v},{\bf u}_x]
             +[{\bf v}-2{\bf u},{\bf v}_x]+[{\bf u}^2,{\bf v}]+2[{\bf u},{\bf v}^2]\bigr),\\
 {\bf v}_t &= -{\bf v}_{xx}+{\bf u}{\bf u}_x+{\bf u}_x{\bf u} 
             -{\bf u}{\bf v}^2+2{\bf v}{\bf u}{\bf v}-{\bf v}^2{\bf u} \\
     &\qquad +i\frac{\sqrt{3}}{3}\bigl([{\bf u}-2{\bf v},{\bf u}_x]
             +2[{\bf u}+{\bf v},{\bf v}_x]+2[{\bf u}^2,{\bf v}]+[{\bf u},{\bf v}^2]\bigr).
\end{aligned}\right.
\end{align}
The scalar system is invariant with respect to the scaling with the cubic root of 1:
\[
 {\bf u}\to\varepsilon {\bf u},\qquad {\bf v}\to\varepsilon^2{\bf v},\qquad \varepsilon=-1/2+i\sqrt{3}/2.
\]
The system \ref{B'4} also does not change under this transformation, while \ref{B''4} turns into two another systems with different coefficients. Applying the involution (\ref{star1}) we obtain in total 8 noncommutative analogs for \ref{B4}.

The linear change
\[
 \partial_T=i\sqrt{3}\,\partial_t,\qquad 
 {\bf p}=\frac{1}{2}({\bf u}-{\bf v}),\qquad 
 {\bf q}=\frac{i}{2\sqrt{3}}({\bf u}+{\bf v})
\]
brings to radical-free systems with a non-standard separant: \ref{B'4} and \ref{B''4} take, respectively, the form
\begin{align}
\label{B'4.pq}
&\left\{\begin{aligned}
 {\bf p}_t &= 3({\bf q}_x -{\bf pq}-{\bf qp})_x
               +[{\bf p},{\bf p}_x-6{\bf q}^2]-3[{\bf q},{\bf q}_x]
              -2({\bf p}^2{\bf q}-2{\bf pqp}+{\bf qp}^2),\\
 {\bf q}_t &= (-{\bf p}_x-{\bf p}^2+3{\bf q}^2)_x 
              +[{\bf p}_x+2{\bf p}^2,{\bf q}]-[{\bf p},{\bf q}_x] 
              +2({\bf pq}^2-2{\bf qpq}+{\bf q}^2{\bf p}),
\end{aligned}\right.\\[2mm]
\label{B''4.pq}
&\left\{\begin{aligned}
 {\bf p}_t &= 3({\bf q}_x -{\bf pq}-{\bf qp})_x
              +3[{\bf p},{\bf p}_x-6{\bf q}^2]-9[{\bf q},{\bf q}_x]
              +6({\bf p}^2{\bf q}-2{\bf pqp}+{\bf qp}^2),\\
 {\bf q}_t &= (-{\bf p}_x-{\bf p}^2+3{\bf q}^2)_x 
               -[5{\bf p}_x+2{\bf p}^2,{\bf q}]-3[{\bf p},{\bf q}_x] 
              -6({\bf pq}^2-2{\bf qpq}+{\bf q}^2{\bf p}).
\end{aligned}\right.
\end{align}
A scalar version of these systems was obtained in \cite{Sokolov_Shabat_1980} from a spectral problem with a third order differential operator in factorized form. This spectral problem can be generalized for noncommuting variables. As the result, we arrive at the zero curvature representation (\ref{ZCR}) for the system (\ref{B'4.pq}), with matrices
\begin{align*}
 & U=\begin{pmatrix}
  {\bf q}+{\bf p} &  {\bf I} & 0 \\
   0  & {\bf q}-{\bf p} & {\bf I} \\
  \lambda{\bf I} & 0 & -2{\bf q} 
 \end{pmatrix},\\
 & V=-3U^2
  +\begin{pmatrix}
  3{\bf q}_x-{\bf p}_x & 0 & 0 \\
   0 & -3{\bf q}_x-{\bf p}_x & 0 \\
   0 & 0 & 2{\bf p}_x 
 \end{pmatrix} +2({\bf p}^2+[{\bf p},{\bf q}]+3{\bf q}^2)I.
\end{align*}
The matrices for (\ref{B''4.pq}) differ from them in diagonal elements:
\[
 \widetilde U= U+2{\bf q}\,I,\qquad
 \widetilde V= V+\operatorname{diag}(0,0,6[{\bf p},{\bf q}])
  -2({\bf p}_x+{\bf p}^2+4[{\bf p},{\bf q}]-3{\bf q}^2)\,I.
\]
 
\section{Symmetries of non-Abelian systems}
\label{s:sym}

\subsection{Linearizable systems}\label{linsys}

It is well-known that higher symmetries exist not only for equations which are integrable by the inverse scattering method, but also for the Burgers type equations which admit linearization by differential substitutions (see e.g. \cite{MikShaSok91}). If we relax our requirements and replace the existence of conservation laws with the existence of symmetries then the list from Section \ref{s:com} is extended by linearizable systems.

\begin{theorem}[\cite{SW}] \label{th:sym} 
{\rm i)} Any system of the form \eqref{kvazgen} which admits a symmetry of the form
\begin{equation}
\label{kvazsym}
 \begin{cases}
u_{\tau}=~u_{xxxx}+f(u,v,u_x,v_x,u_{xx},v_{xx},u_{xxx},v_{xxx}), \\[1.5mm]
v_{\tau}= -v_{xxxx}+g(u,v,u_x,v_x,u_{xx},v_{xx},u_{xxx},v_{xxx})
\end{cases}
\end{equation}
is polynomial.

{\rm ii)} A homogeneous polynomial non-triangular system \eqref{kvazgen} admits a symmetry \eqref{kvazsym} if and only if it belongs to the lists from Theorem \ref{th:com} or coincides with one of the systems listed below, up to a scaling of the variables $x,t,u,v$ and the interchange $(u,v)\mapsto(v,u)$: 
\begin{align}
%---------------------------------------------
\label{L1}\tag*{\text{$L_1$}}
&\left\{\begin{aligned}
 u_t &=~~u_{xx }+2uu_{x}+2vu_{x}+2uv_{x}+2u^2v+2uv^2,\\
 v_t &= -v_{xx }-2vv_{x}-2vu_{x}-2uv_{x}-2u^2v-2uv^2,
\end{aligned}\right.\\
%---------------------------------------------
\label{L2}\tag*{\text{$L_2$}}
&\left\{\begin{aligned}
 u_t &=~~u_{xx }+2u_{x}v+2uv_{x}+2uv^2+u^2,\\
 v_t &= -v_{xx }-2vv_{x}-u_{x},
\end{aligned}\right.\\
%---------------------------------------------
\label{L3}\tag*{\text{$L_3$}}
&\left\{\begin{aligned}
 u_t &=~~u_{xx }+2\alpha u^2v_{x}+2\alpha uvu_{x}-\alpha\beta u^3v^2,\\
 v_t &= -v_{xx }+2\beta v^2u_{x}+2\beta uvv_{x}+\alpha\beta u^2v^3,
\end{aligned}\right. \qquad \alpha\beta\ne0, \\
%---------------------------------------------
\label{L4}\tag*{\text{$L_4$}}
&\left\{\begin{aligned}
 u_t &=~~u_{xx }+4uvu_{x}+4u^2v_{x}+3vv_{x}+2u^3v^2+uv^3,\\
 v_t &= -v_{xx }-2v^2u_{x}-2uvv_{x}-2u^2v^3-v^4,
\end{aligned}\right.\\
%---------------------------------------------
\label{L5}\tag*{\text{$L_5$}}
&\left\{\begin{aligned}
 u_t &=~~u_{xx }+4uu_{x}+2vv_{x},\\
 v_t &= -v_{xx }-2vu_{x}-2uv_{x}-3u^2v-v^3.
\end{aligned}\right.
\end{align}
\end{theorem}

In order to obtain an alternative verification of the results presented in Theorem \ref{th:noncom} and to find noncommutative analogs for the systems \ref{L1}--\ref{L5} we use a criterion based on the existence of symmetries.

\begin{theorem}\label{th:noncom'} 
All non-equivalent non-Abelian analogs of the systems \ref{S1}--\ref{S5}, \ref{B1}--\ref{B4} and \ref{L1}--\ref{L5}, which satisfy the following assumptions:
\begin{itemize}
\item[---] the non-Abelian analog admits a symmetry of the same minimal order as the original system;
\item[---] the non-Abelian analog and its symmetry are polynomial, homo\-geneous and admit the scaling group
\[
 (x,\ t,\, {\bf u},\, {\bf v})\longrightarrow (\tau^{-1}x,\ \tau^{-\mu} t,\, \tau^{\nu_1} {\bf u}, \tau^{\nu_2} {\bf v})
\]
with the same $\mu$, $\nu_1$ and $\nu_2$ as for the original system and its symmetry;
\item[---] the non-Abelian analog and its symmetry turn into the original system and its symmetry under substitution of commuting variables instead of noncommuting ones {\rm(}or, in the matrix language, for the $1\times1$ matrices{\rm)},
\end{itemize}
are exhausted by the systems from Theorem \ref{th:noncom} and the systems \ref{L'1}, \ref{L'2}, \ref{L'310} and \ref{L''310} given below. 
\end{theorem}

The system \ref{L1} has, up to the involution, one noncommutative generalization
\begin{align}\label{L'1}\tag*{\text{$L'_1$}}
\left\{\begin{aligned}
 {\bf u}_t &=~~{\bf u}_{xx}+2{\bf u}_x{\bf u}+2({\bf uv})_x+2{\bf uv}({\bf u}+{\bf v}),\\
 {\bf v}_t &= -{\bf v}_{xx}-2{\bf v}_x{\bf v}-2({\bf vu})_x-2{\bf vu}({\bf u}+{\bf v}).
\end{aligned}\right.
\end{align}
It is related with the system $\tilde{\bf u}_t= \tilde{\bf u}_{xx}$, $\tilde{\bf v}_t = -\tilde{\bf v}_{xx}$ by the differential substitution
\[
 {\bf u}= \tilde{\bf u}_x (\tilde{\bf u}+\tilde{\bf v})^{-1},\qquad 
 {\bf v}= \tilde{\bf v}_x (\tilde{\bf u}+\tilde{\bf v})^{-1}.
\]
Here we assume that the algebra ${\cal A}$ is extended by additional element $({\bf u}+{\bf v})^{-1}$. 

The system \ref{L2} also has one noncommutative analog,
\begin{align}\label{L'2}\tag*{\text{$L'_2$}}
\left\{\begin{aligned}
 {\bf u}_t &=~~{\bf u}_{xx}+2({\bf uv})_x+{\bf u}^2+2{\bf uv}^2,\\
 {\bf v}_t &= -{\bf v}_{xx}-{\bf u}_x-2{\bf v}_x{\bf v}-[{\bf u},\,{\bf v}],
\end{aligned}\right.
\end{align}
which is linearized by substitution
\[
 {\bf u}= \tilde{\bf u}_x\tilde{\bf u}^{-1},\qquad 
 {\bf v}= \tilde{\bf v}_x\tilde{\bf u}^{-1},\qquad \mbox{where} \qquad
 \tilde{\bf u}_t= \tilde{\bf u}_{xx},~~ 
 \tilde{\bf v}_t= -\tilde{\bf v}_{xx}-\tilde{\bf u}_x.
\]

For the family $L_3(\alpha,\beta)$ of the Eckhaus equation type \cite{Calogero_Eckhaus_1987, Calogero_Degasperis_Lillo} one can assume, taking the scaling into account, that $\alpha=1$ and $\beta$ is a free parameter. It turns out that non-Abelian generalizations exist only for $\beta=0$, that is, when the scalar system becomes triangular. Although this degenerate system is not considered in Theorem \ref{th:sym}, it still has higher symmetries and can be treated in the same manner as other examples. For this system there exist two non-Abelian analogs (up to the involution), moreover, one of them is not triangular:
\begin{align}
\label{L'310}\tag*{\text{$L'_3(1,0)$}}
&\left\{\begin{aligned}
 {\bf u}_t &=~~{\bf u}_{xx}+2{\bf u}{\bf v}_x{\bf u}+2{\bf u}_x{\bf v}{\bf u},\\
 {\bf v}_t &= -{\bf v}_{xx},
\end{aligned}\right.\\[1.5mm]
\label{L''310}\tag*{\text{$L''_3(1,0)$}}
&\left\{\begin{aligned}
 {\bf u}_t &=~~{\bf u}_{xx}+2{\bf u}_x{\bf uv}+2{\bf u}^2{\bf v}_x
              -2{\bf u}^2[{\bf u},{\bf v}]{\bf v},\\
 {\bf v}_t &= -{\bf v}_{xx}+2[{\bf uv},{\bf v}_x-{\bf uv}^2]
              +2[{\bf u}_x{\bf v}+{\bf u}^2{\bf v}^2,{\bf v}].
\end{aligned}\right.
\end{align}
In these cases, the linearization is carried out as a composition of counter-directional differential substitutions into some intermediate system from the given and from a linear ones. The system \ref{L'310} admits the following substitution $({\bf u},{\bf v})\to({\bf w},{\bf v})$:   
\[
 {\bf w}={\bf u}_x{\bf u}^{-1}+{\bf uv},\qquad \mbox{where} \qquad 
 {\bf w}_t= {\bf w}_{xx}+2{\bf w}_x{\bf w}, \qquad {\bf v}_t = -{\bf v}_{xx},
\]
and the Burgers equation for ${\bf w}$ is linearized by the non-Abelian Cole--Hopf substitution ${\bf w}=\tilde{\bf u}_x\tilde{\bf u}^{-1}$, where $\tilde{\bf u}_t= \tilde{\bf u}_{xx}$. This define the transform $({\bf u},{\bf v}) \leftrightarrow(\tilde{\bf u},{\bf v})$ which is implicit in both directions (B\"acklund type transformation or correspondence). Similarly, the system \ref{L''310} admits the substitution
\[
 {\bf w}={\bf u}_x{\bf u}^{-1}+{\bf u}^2{\bf vu}^{-1},\qquad
 {\bf p}={\bf uvu}^{-1},
\]
which brings to the system
\[
 {\bf w}_t= {\bf w}_{xx}+2{\bf w}_x{\bf w},\qquad 
 {\bf p}_t = -{\bf p}_{xx}+2[{\bf w},{\bf p}]_x+2[{\bf w},{\bf p}]{\bf w},
\]
and the latter is linearized by the substitution
\[
  {\bf w}=\tilde{\bf u}_x\tilde{\bf u}^{-1},\qquad
  {\bf p}=\tilde{\bf u}\tilde{\bf v}\tilde{\bf u}^{-1},\qquad \mbox{where} \quad 
  \tilde{\bf u}_t= \tilde{\bf u}_{xx},\quad 
  \tilde{\bf v}_t= -\tilde{\bf v}_{xx}. 
\]
The systems $L_3(1,\beta)$ with $\beta\ne0$, \ref{L4} and \ref{L5} have no noncommutative generalizations admitting the higher symmetries.

\subsection{Examples of master-symmetries and B\"acklund transformations}

Some of non-Abelian systems presented in the paper admit master-symmetries or B\"acklund transformations in the form of integrable lattice equations. We have not investigated these additional structures in full generality and will only give a few examples that we have been able to find.

For the sake of simplicity we restricted ourselves by consideration of local master-symmetries, although they usually involve nonlocalities even in the scalar case. For instance, a local master-symmetry exists for the system \ref{S2} (see e.g.~the review article \cite{ASY} where master-symmetries were given for some of NLS type systems). We have found a generalization only for \ref{S'2}:
\begin{equation}\label{S'2.ms}
\left\{\begin{aligned}
 {\bf u}_\tau &= x({\bf u}_{xx}+2{\bf u}_x{\bf u}+2({\bf vu})_x+2[{\bf vu},{\bf u}])
  +2{\bf u}_x+{\bf u}^2+3{\bf vu},\\
 {\bf v}_\tau &= x(-{\bf v}_{xx}+2{\bf v}_x{\bf v}+2({\bf uv})_x+2[{\bf v},{\bf uv}])
  -2{\bf v}_x+{\bf v}^2+3{\bf uv}.
\end{aligned}\right.
\end{equation}
A master-symmetry for the system \ref{S''2} remains unknown (it may exist, but it may require the introduction of some nonlocality, which disappears in the scalar case). Another well-known result is that the system \ref{S2} admits the B\"acklund-Schlesinger transformation in the form of the Volterra lattice \cite{Levi_1981}. A noncommutative generalization of this fact is known for the system \ref{S''2}: one can prove by a direct computation that the non-Abelian Volterra lattice \cite{Salle_1982}
\begin{equation}\label{VL.x}
 {\bf q}_{n,x}={\bf q}_n{\bf q}_{n+1}-{\bf q}_{n-1}{\bf q}_n
\end{equation}
possesses the symmetry
\begin{equation}\label{VL.t}
 {\bf q}_{n,t}={\bf q}_n{\bf q}_{n+1}{\bf q}_{n+2} +{\bf q}_n{\bf q}^2_{n+1}+{\bf q}^2_n{\bf q}_{n+1} -{\bf q}_{n-1}{\bf q}^2_n -{\bf q}^2_{n-1}{\bf q}_n-{\bf q}_{n-2}{\bf q}_{n-1}{\bf q}_n
\end{equation}
and that for any $n$ the variables ${\bf u}={\bf q}_{n+1}$, ${\bf v}={\bf q}_n$ satisfy the system \ref{S''2} in virtue of these two lattice equations.

The system \ref{S'512} has the local master-symmetry
\begin{equation}\label{S'512.ms}
\left\{\begin{aligned}
 {\bf u}_t &= x{\bf u}_{xx}+2(x{\bf u}{\bf v}{\bf u})_x,\\
 {\bf v}_t &= -x{\bf v}_{xx}+2(x{\bf v}{\bf u}{\bf v})_x-3{\bf v}_x.
\end{aligned}\right.
\end{equation}
The B\"acklund-Schlesinger transformation for the DNLS-I systems is defined by the modified Volterra lattice. It is known that it admits two non-Abelian generalizations. We write down both the lattice and its symmetry of second order. The pair of equations 
\begin{align*}
 {\bf q}_{n,x}&={\bf q}_n({\bf q}_{n+1}-{\bf q}_{n-1}){\bf q}_n, \\[1.5mm]
 {\bf q}_{n,t}&={\bf q}_n{\bf q}_{n+1}({\bf q}_{n+2}+{\bf q}_n){\bf q}_{n+1}{\bf q}_n
  -{\bf q}_n{\bf q}_{n-1}({\bf q}_n+{\bf q}_{n-2}){\bf q}_{n-1}{\bf q}_n
\end{align*}
is consistent and the variables ${\bf u}={\bf q}_{n+1}$, ${\bf v}={\bf q}_n$ satisfy the system \ref{S'512} \cite{Adler_Svinolupov_Yamilov_1999}. Similarly, for the consistent pair
\begin{align*}
 &{\bf q}_{n,x}={\bf q}_{n+1}{\bf q}^2_n-{\bf q}^2_n{\bf q}_{n-1}, \\[2mm]
 &\begin{aligned}
 {\bf q}_{n,t}&={\bf q}_{n+2}{\bf q}^2_{n+1}{\bf q}^2_n 
        +{\bf q}_{n+1}{\bf q}_n{\bf q}_{n+1}{\bf q}^2_n 
        +{\bf q}_{n+1}{\bf q}^2_n{\bf q}_{n-1}{\bf q}_n\\ 
        &\qquad -{\bf q}_n{\bf q}_{n+1}{\bf q}^2_n{\bf q}_{n-1} 
        -{\bf q}^2_n{\bf q}_{n-1}{\bf q}_n{\bf q}_{n-1} 
        -{\bf q}^2_n{\bf q}^2_{n-1}{\bf q}_{n-2},
\end{aligned}
\end{align*}
the variables ${\bf u}={\bf q}_{n+1}$ and ${\bf v}={\bf q}_n$ satisfy \ref{S''512} (notice that the substitution ${\bf\tilde q}_n={\bf q}_{n+1}{\bf q}_n$ leads to the Volterra lattice (\ref{VL.x}), up to the involution). 

It would be interesting to find lattice representations for other non-Abelian systems. However, such lattice equations can be not polynomial even in the scalar case, and their generalization may require consideration of noncommutative Laurent polynomials.

\section{Conclusion and perspectives}

In this paper, we have found non-Abelian generalizations for some key integrable systems of NLS and Boussinesq types. Basically, we restricted ourselves to the case of homogeneous polynomial quasilinear systems of the form (\ref{kvazgen}). For this class of equations we were able to give complete classifications of non-Abelian analogs based both on the existence of higher conservation laws (Theorem \ref{th:noncom}) and the higher symmetries (Theorem \ref{th:noncom'}). While some of the generalizations obtained are well-known, there were also surprisingly many new examples. The integrability of each equation is justified either by explicit zero-curvature representation or by linearizing substitution.

Our approach is suitable not only for quasilinear systems. The reason for this restriction was only that all such systems are polynomial and their complete list was already obtained in \cite{SW}. It would be quite possible to find noncommutative generalizations for any integrable polynomial homogeneous systems \eqref{sys2} with positive weights, however one should start this project by compiling a complete list of such scalar systems. 

\begin{example}
The well-known scalar system \cite{MikShaYam87}
\begin{equation}\label{B5}
 \left\{\begin{aligned}
 &u_t=u_{xx}+6v^2_x+18v^2v_x-6uv_x,\\
 &v_t=-v_{xx}+u_x,
 \end{aligned}\right.
\end{equation}
which contain a quadratic term in derivatives, has the following non-Abelian analogs:
\begin{align}
\label{B'5}
&\left\{\begin{aligned}
 {\bf u}_t &=~~{\bf u}_{xx}+6{\bf v}^2_x+6({\bf v}^3)_x-3{\bf uv}_x-3{\bf v}_x{\bf u}\\
    &\qquad +i\sqrt{3}\bigl(2[{\bf v},\,{\bf v}_{xx}]+3[{\bf v}^2,\,{\bf v}_x-{\bf u}]
                +[{\bf u},\,{\bf v}]_x\bigr),\\[1.4mm]
 {\bf v}_t &= -{\bf v}_{xx}+{\bf u}_x+i\sqrt{3}[{\bf v},\,{\bf v}_x-{\bf u}],
\end{aligned}\right.
\end{align}
and
\begin{align}
\label{B''5}
&\left\{\begin{aligned}
 {\bf u}_t &=~~{\bf u}_{xx}+6{\bf v}^2_x+18{\bf vv}_x{\bf v}-3{\bf uv}_x-3{\bf v}_x{\bf u}\\
    &\qquad -i\sqrt{3}\bigl(2[{\bf v},\,{\bf v}_{xx}]+3[{\bf v}^2,\,{\bf v}_x-{\bf u}]
                -[{\bf u},\,{\bf v}]_x\bigr),\\[1.4mm]
 {\bf v}_t &= -{\bf v}_{xx}+{\bf u}_x+i\sqrt{3}[{\bf v},\,{\bf v}_x-{\bf u}].
\end{aligned}\right.
\end{align}
The differential substitution
\[
 {\bf\tilde u}=2{\bf u}_x+3({\bf u}+{\bf v})^2,\qquad {\bf\tilde v}={\bf u}+{\bf v}
\]
relates the systems \eqref{B'5} (written for the variables with tilde) with the system \ref{B'3}.
\end{example}

As already noted, some of the systems from Theorem \ref{th:com} admit the addition of terms of lower weight preserving the integrability property. Non-commutative generalizations may exist for such systems as well.

\begin{example} 
It is well-known that the scalar system \ref*{S5}$(1,2)$ (DNLS-I) admits the following generalization by adding the linear terms \cite{MikShaYam87,SW}:
\begin{equation}\label{S512.ab}
\left\{\begin{aligned}
 u_t &=~~u_{xx}+2(u^2v)_x+\alpha v_x,\\
 v_t &= -v_{xx}+2(uv^2)_x+\beta u_x
\end{aligned}\right.
\end{equation}
(the constants $\alpha$ and $\beta$ can be scaled either to 1 or 0, without loss of generality, so that, taking the interchange of $u$ and $v$ into account, we have here just two essentially different versions in addition to the homogeneous case). There exist two noncommutative analogs for this system which coincide with \ref{S'512} and \ref{S''512} for the zero values of parameters. The non-homogeneous extension of the system \ref{S'512} is of the form
\begin{equation}
\label{S'512.ab}
\left\{\begin{aligned}
 {\bf u}_t &=~~{\bf u}_{xx}+2({\bf uvu})_x-2\gamma^2{\bf v}_x
              -2\gamma[{\bf u},{\bf v}]_x
              -2\delta[{\bf u},{\bf u}_x+2{\bf uvu}] \\
           &\qquad +4\gamma\delta({\bf u}^2{\bf v}-2{\bf uvu}+{\bf vu}^2)
              +4\gamma^2\delta[{\bf u},{\bf v}],\\[1.5mm]
 {\bf v}_t &= -{\bf v}_{xx}+2({\bf vuv})_x-2\delta^2{\bf u}_x
               +2\delta[{\bf u},{\bf v}]_x
               -2\gamma[{\bf v},{\bf v}_x-2{\bf vuv}] \\
            &\qquad -4\gamma\delta({\bf uv}^2-2{\bf vuv}+{\bf v}^2{\bf u})
               +4\gamma\delta^2[{\bf u},{\bf v}],
\end{aligned}\right.
\end{equation}
where $\alpha=-2\gamma^2$ and $\beta=-2\delta^2$. The signs of $\gamma$ and $\delta$ can be chosen arbitrarily, resulting in four extensions of the scalar system (as usual, the involution halves this number). For the non-Abelian system \ref{S''512}, the admissible terms of lower weights are simpler:
\begin{equation}\label{S''512.ab}
\left\{\begin{aligned}
 {\bf u}_t &=~~{\bf u}_{xx}+2{\bf u}^2{\bf v}_x+2{\bf u}_x{\bf u}{\bf v}
     +2{\bf u}_x{\bf v}{\bf u}+2[{\bf u}^2{\bf v}^2,{\bf u}]
     +\alpha({\bf v}_x+[{\bf v}^2,{\bf u}]), \\
 {\bf v}_t &= -{\bf v}_{xx}+2{\bf u}_x{\bf v}^2+2{\bf u}{\bf v}{\bf v}_x
     +2{\bf v}{\bf u}{\bf v}_x -2[{\bf v},{\bf u}^2{\bf v}^2]
     +\beta({\bf u}_x+[{\bf u}^2,{\bf v}]). \\
\end{aligned}\right.
\end{equation}
The statement is that the systems (\ref{S'512.ab}) and (\ref{S''512.ab}) give all non-Abelian generalizations of (\ref{S512.ab}), up to the involution, and that both systems admit higher symmetries and conservation laws for any values of parameters.
\end{example}

A limitation of our approach is the assumption that the weights of $u$ and $v$ should be positive. This is always true for quasilinear systems, but there exist also integrable equations of more general form for which the weights may be zero or even negative. A simple example is given by the potential DNLS system 
\[
 \left\{\begin{aligned}
  & u_t=~~u_{xx}+2u^2_xv_x,\\     
  & v_t= -v_{xx}+2u_xv^2_x
 \end{aligned}\right.
\]
with $\nu_1+\nu_2=-1$. In such cases, it is necessary to introduce additional selection rules in order to bound the number of monomials under consideration.

The setting of classification problems for the noncommutative systems also requires pondering. The main problem is that the set of admissible transformations becomes significantly narrower compared to the commutative case. For instance, it was proved in \cite{Mikhailov_Shabat_1985} that if the separant of a scalar integrable system depends on $u$ and $v$ then this system can be brought to the form \eqref{sys2} by some change of variables. The examples \eqref{B'5} and \eqref{B''5} suggest that this may be not true in for non-Abelian systems. Because of this, the scalar systems related by point changes (not saying about more general transformations) with systems from Theorem \ref{th:com} may have noncommutative generalizations which are not equivalent to the systems obtained in our paper.

Although it is clear that the theory of transformations must be changed in the noncommutative setting, at the moment it is difficult to say how this affects the choice of the canonical forms of integrable equations. The examples related with systems of the Boussinesq type suggest that in some aspects other canonical forms may turn more convenient.

The generalization of the method to various classes of systems with rational right-hand side also requires additional research. An example of such a system is the non-Abelian Heisenberg equation \cite{SviSok94}
\[
 \left\{\begin{aligned}
 {\bf u}_t&=~~{\bf u}_{xx} -2\,{\bf u}_x ({\bf u}+{\bf v})^{-1} {\bf u}_{x},\\
 {\bf v}_t&= -{\bf v}_{xx} +2\,{\bf v}_x ({\bf u}+{\bf v})^{-1} {\bf v}_{x}.
\end{aligned}\right. 
\]

Thus, we see that in the theory of noncommutative integrable equations, there are many unsolved problems, with potentially rich and interesting answers. Their study should be the subject of further research.

%-------------------------------------------------------------------------------
\subsubsection*{Acknowledgements}

This work was carried out under the State Assignment 0033-2019-0006 (Integrable systems of mathematical physics) of the Ministry of Science and Higher Education of the Russian Federation.
 
%-------------------------------------------------------------------------------

\end{document}